\documentclass[11pt]{article}

\usepackage[preprint]{acl}

\usepackage{times}
\usepackage{latexsym}
\usepackage[T1]{fontenc}
\usepackage[utf8]{inputenc}
\usepackage{microtype}
\usepackage{inconsolata}
\usepackage{graphicx}
\usepackage{booktabs}
\usepackage{amsmath}
\usepackage{xcolor}
\usepackage{xspace}
\usepackage{enumitem}
\usepackage[most]{tcolorbox}
\usepackage{fvextra}

\usepackage{pifont}
\newcommand{\cmark}{\ding{51}}%
\newcommand{\xmark}{\ding{55}}%

\newcommand{\simdomain}{CAE simulation\xspace}

\title{What Do CAE Simulation Agents Really Need Beyond a Generic Harness?}

\author{Jiasheng Shi \\
  DP Technology \\
  Beijing, China
\And
  Tianhan Zhang\thanks{\;Corresponding author.} \\
  School of Astronautics, Beihang University \\
  AI for Science Institute \\
  Beijing, China \\
  \texttt{thzhang@buaa.edu.cn}}

\begin{document}
\maketitle

\begin{abstract}
Computer-aided engineering (CAE) simulation is among the largest and most demanding areas of engineering, where setting up a solver such as OpenFOAM, FEniCS, or COMSOL takes real expertise. Large language model (LLM) agents promise to turn a natural-language request into a working simulation, and recent CAE agents add simulation-specific machinery: multi-agent decomposition, domain retrieval, and scripted reflection. That machinery suited weak base models; modern harnesses already supply multi-turn reasoning, tool use, and execution feedback. We ask what a CAE simulation agent still needs beyond a generic harness. With information access and repair budget held fixed, a single-agent harness matches or beats multi-agent specialized systems (FoamBench 96.4\% vs.\ 88.2\%).
 Ablations trace this to capabilities the harness already provides: execution-feedback repair lifts FoamBench from 71.8\% with no repair round to 96.4\%, while scripted reflection adds nothing. The one input that still helps is domain knowledge supplied as solver tutorials, our largest measured gain (80.9\% to 96.4\%).

\end{abstract}

\section{Introduction}
\label{sec:intro}
\begin{figure*}[t]
\centering
\includegraphics[width=1.0\linewidth]{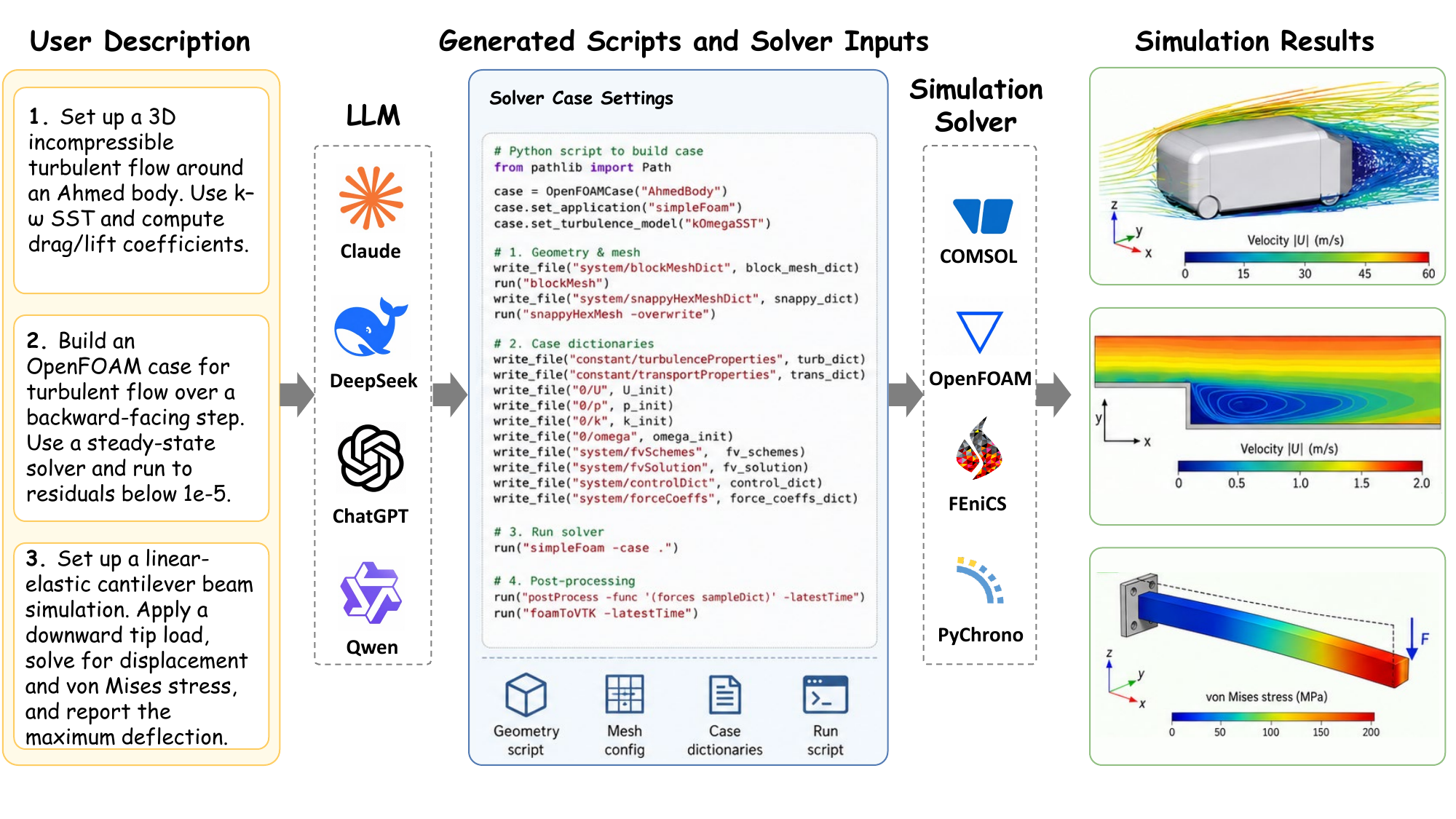}
\caption{Overview of a \simdomain task. A user provides a natural-language description of the simulation scenario (e.g., turbulent flow around an Ahmed body); an LLM agent generates the necessary scripts, case dictionaries, and solver configurations; simulation backends (OpenFOAM, COMSOL, FEniCS, etc.) execute the case and produce velocity fields, pressure distributions, convergence residuals, and integrated quantities such as drag and lift coefficients.}
\label{fig:overview}
\end{figure*}

\simdomain, spanning computational fluid dynamics~\cite{anderson2013computational}, finite element analysis~\cite{kim2026introduction}, and multiphysics modeling, is one of the largest and most demanding areas of modern engineering, central to design in aerospace, automotive, civil, and energy applications.
Running a case correctly is hard: it relies on solvers such as OpenFOAM~\cite{greenshields2022} or COMSOL~\cite{comsol2025}, and on trained engineers who understand the governing equations, boundary conditions, meshing, solver settings, and numerical schemes, which puts these tools out of reach for users without that background.
Large language model (LLM) agents promise to lower this barrier: the user writes a short natural-language description, and the agent runs the simulation toolchain end to end on the user's behalf (Figure~\ref{fig:overview}).

Recent CAE agents pursue this with a common recipe: multi-agent role decomposition, domain-specific retrieval-augmented generation (RAG), and \emph{scripted reflection} (throughout, we use this term for explicit self-critique instructions added to the prompt, a pattern inherited from reviewer-style scaffolding), tightly coupled to a particular solver.
Systems such as Foam-Agent~\cite{abs-2505-04997, abs-2509-18178}, MetaOpenFOAM~\cite{abs-2407-21320, abs-2502-00498}, and ChatCFD~\cite{abs-2506-02019} report strong results under this design.
This complexity was a rational response to earlier LLMs, which could not reliably read governing equations or solver syntax and planned poorly under short context windows and fragile tool use: role decomposition compensated for weak planning, and curated retrieval for missing domain knowledge.

The agent landscape has since shifted.
A new generation of general-purpose coding-agent harnesses, such as Claude Code and Codex, combines long-horizon multi-turn context, file and shell tools, sub-agent dispatch, and tight execution feedback on top of stronger base models~\cite{abs-2604-02460,abs-2601-12307,xiao2026preliminary,abs-2604-14228}.
Evidence from benchmarks such as RExBench~\cite{abs-2506-22598} shows that much of what specialized scaffolds once added, including structured planning, repeated repair, and knowledge lookup, is now performed by a generic harness with no domain-specific code.
In other words, multi-turn reasoning and iterative repair are no longer capabilities a CAE agent must build; they come with the harness.

This motivates a concrete question: what does a CAE simulation agent still need beyond a generic harness?
We do not argue that prior specialized systems were misguided; in their respective base-model eras, the additional structure carried real weight.
What has changed is its \emph{marginal} return on top of a modern general-purpose harness, which current benchmark headline numbers do not cleanly isolate.

We study this question empirically.
We adopt an intentionally minimal Direct Baseline: a strong general-purpose LLM driven by an off-the-shelf coding-agent harness that exposes multi-turn context, file and shell tools, sub-agent dispatch, and execution feedback, with no simulation-specific scaffolding added, that is, no solver-oriented role decomposition, no curated domain RAG, no scripted-reflection module, and no solver-specific orchestration.
We compare the Direct Baseline against representative specialized agents on CAE tasks spanning OpenFOAM, FEniCS, PyChrono, and COMSOL, holding information access, repair budget, and success criteria fixed.
On top of this comparison we run mechanism ablations that separate the capabilities a harness already provides, namely knowledge access, iterative repair driven by execution feedback, and multi-turn reasoning, from the simulation-specific scaffolds layered on top.

Our findings are constructive rather than dismissive:
\begin{itemize}[leftmargin=*]
\item Under matched conditions, a single-agent Direct Baseline matches or beats multi-agent specialized systems; the capabilities driving this performance are the harness's own, while scripted reflection and role decomposition add little on top.
\item The one input that still requires human engineering is domain knowledge, supplied simply as solver tutorials, and it is the largest gain we measure. Multi-turn reasoning and iterative repair are decisive but now built into the harness, so engineer effort is best spent encoding domain expertise rather than assembling scaffolds.
\item Progress is limited by evaluation: most CAE benchmarks check only whether generated code runs, not whether the result is physically correct or representative of industrial complexity. We argue the field's main need is standardized, representative benchmarks through which engineers can share domain expertise and that genuinely discriminate among agents at industrial complexity.
\end{itemize}

\section{Related Work}
\label{sec:related}

\subsection{Agentic Tool}

Tool-using LLM agents have progressed through three increasingly
capable stages.
Early work treated the LLM itself as a planning agent that
decomposes tasks into natural-language sub-goals.
A second stage introduced \emph{code agents} that translate plans
into executable programs and iterate on
errors~\cite{abs-2409-02977}.
The current stage couples LLMs with general-purpose
harnesses, file editing, shell execution, and structured tool
invocation, exemplified by Claude Code and analogous
scaffolds~\cite{abs-2604-14228,abs-2512-10398}
and probed by demanding benchmarks such as
RExBench~\cite{abs-2506-22598}.
A parallel line of
evidence~\cite{abs-2604-02460,abs-2601-12307,xiao2026preliminary}
further shows that a single strong agent driven by such a harness
already matches or surpasses elaborate multi-agent systems on
coding and CFD tasks, raising the question whether specialized
simulation agents still offer marginal value.

\subsection{Specialized Simulation Agents and Benchmarks}

Specialized simulation systems converge on three recurring scaffolding patterns.
\emph{Role decomposition} coordinates planner, writer, runner, and reviewer agents around a target solver: OpenFOAM in Foam-Agent~\cite{abs-2505-04997,abs-2509-18178},
MetaOpenFOAM~\cite{abs-2407-21320,abs-2502-00498,abs-2503-01273}, and
ChatCFD~\cite{abs-2506-02019}; FEniCS in
MCP-SIM~\cite{park2026self}; OpenSeesPy in
MASSE~\cite{abs-2510-11004}; COMSOL in
FEABench~\cite{abs-2504-06260}; with the same recipe extended to a
range of other engineering and scientific
domains~\cite{feng2026openfoamgpt,abs-2510-21993,abs-2504-09754,abs-2311-08166,abs-2601-05187,abs-2511-07262,abs-2512-19458,LiuAGL26}.
\emph{Knowledge injection} via retrieval or fine-tuning is the
second pattern, with ChatCFD, ALL-FEM~\cite{abs-2603-21011}, and
NL2FOAM~\cite{abs-2504-09602} attributing large fractions of their
reported gains to a curated corpus.
\emph{Scripted reflection}, i.e.\ adding explicit self-reflect or reviewer instructions into the prompt so that the agent critiques its own output before committing, forms the third recurring pattern, routinely credited as the most impactful module in published ablations.
A complementary line of work releases solver-coverage benchmarks
without a bespoke agent, including SimBench~\cite{abs-2408-11987} for
PyChrono, FEM-Bench~\cite{abs-2512-20732} for FEniCS, and
CFDCodeBench~\cite{abs-2509-18178} for from-scratch PDE solvers.
Appendix~\ref{app:benchmarks} consolidates the benchmarks
(Table~\ref{tab:benchmarks}) and the specialized systems
(Table~\ref{tab:systems}) covered in this study.

These design choices were reasonable engineering responses to the capabilities of earlier LLMs, and have produced valuable insights on role orchestration, knowledge injection, and iterative refinement. Our work does not dismiss these directions, but asks how much of their marginal value survives once a general-purpose coding harness already provides multi-turn reasoning, execution feedback, and tool use.

\section{Methodology}
\label{sec:method}

\begin{figure}[t]
\centering
\includegraphics[width=1.0\linewidth]{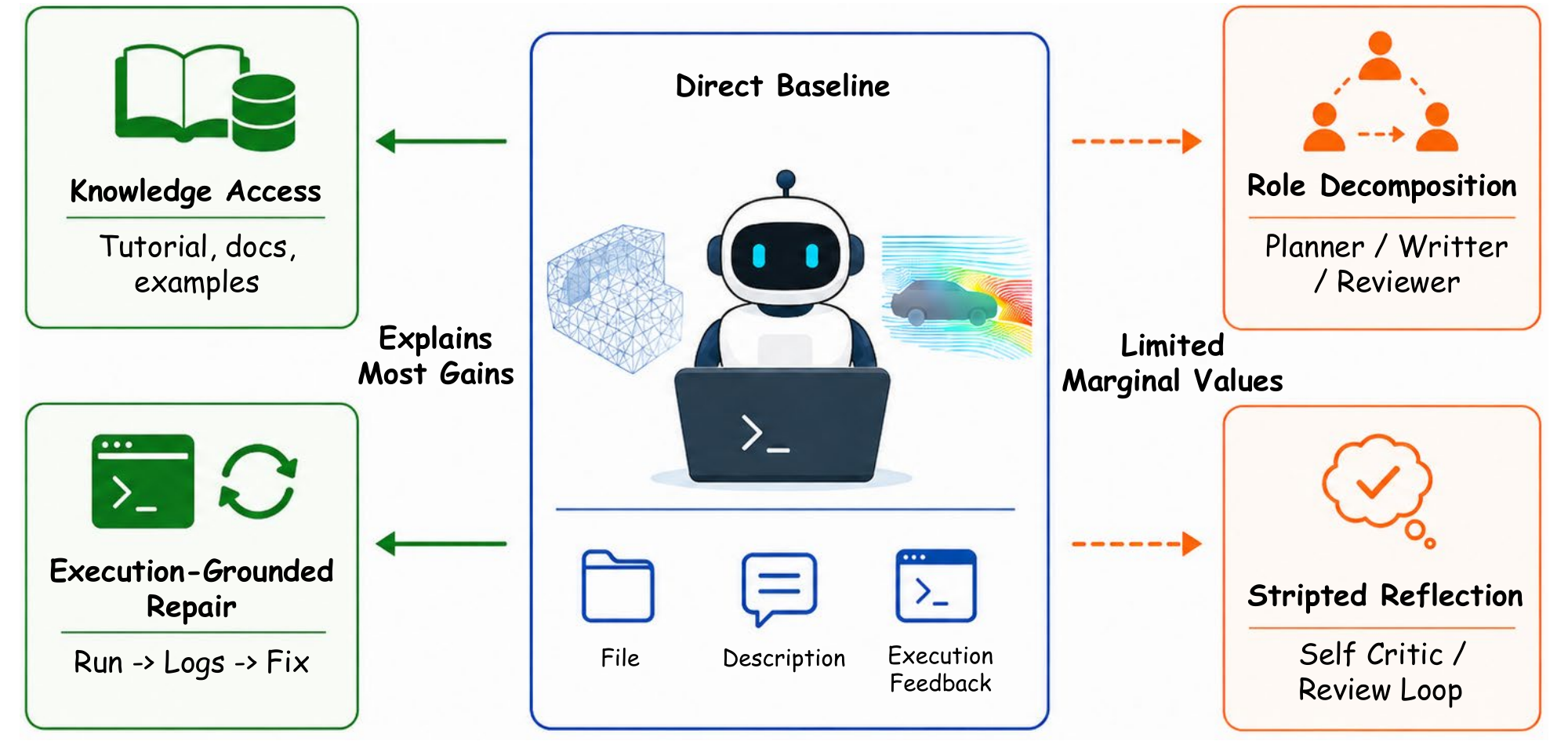}
\caption{Capability attribution framework for \simdomain agents.}
\label{fig:framework}
\centering
\end{figure}

% A generic agent harness already provides the core execution loop, while knowledge access and execution-grounded repair explain most observed gains; role decomposition and scripted reflection add limited marginal value on top.

\subsection{Framework}

Our methodology is built around a minimal
Direct Baseline: rather than building a simulation-specific
orchestrator, we drive each \simdomain task end to end
with an \emph{off-the-shelf, general-purpose coding-agent harness}.
A harness here means a publicly available agent loop that already
exposes file editing, shell execution, and structured tool invocation
on top of a strong LLM, without any simulation-specific scaffolding.
To rule out the possibility that our conclusions hinge on a single
provider, we instantiate the Direct Baseline with four such
harnesses spanning three ecosystems:
(i) Claude Code\footnote{\url{https://claude.com/claude-code}} driving
Claude Opus 4.6\footnote{\url{https://www.anthropic.com/claude/opus}};
(ii) Codex CLI\footnote{\url{https://github.com/openai/codex}} driving
GPT-5.5\footnote{\url{https://openai.com/index/introducing-gpt-5-5/}};
(iii) CodeWhale (DeepSeek-TUI)\footnote{\url{https://github.com/Hmbown/CodeWhale}} driving
DeepSeek V4\footnote{\url{https://api-docs.deepseek.com/}}; and
(iv) opencode\footnote{\url{https://opencode.ai}} driving
Qwen3.5-Plus\footnote{\url{https://qwen.ai}}.
We refer to these four configurations hereafter by their backbone
model -- Claude Opus 4.6, GPT-5.5, DeepSeek V4, and Qwen3.5-Plus --
and use them interchangeably under matched conditions throughout
the study.

Our evaluation isolates two classes of agent capabilities.
\emph{General-purpose capabilities} include multi-turn
reasoning, execution feedback (appending logs after each run),
knowledge access (reading tutorials and API documentation), and
iterative repair (fixing errors based on observed failures).
Modern coding-agent harnesses already expose these through their
core primitives: file/shell tools, multi-turn context windows, and
sub-agent dispatch.
\emph{Domain-specific scaffolds} include solver-oriented role
decomposition (splitting planner, writer, runner, reviewer into
separate agents), curated domain RAG (retrieval over
simulation-specific corpora), scripted reflection (explicit
self-critique instructions inserted into the prompt, inherited from
specialized reviewer scaffolding), and solver-specific orchestration
(hand-coded coordination logic for a particular simulator).
Prior specialized-agent systems layer domain-specific scaffolds on
top of general-purpose capabilities; our mechanism ablations
(Section~\ref{sec:results}) measure the marginal contribution of the
former when the latter are already in place.

The rest of this section defines the Direct Baseline precisely
(\S\ref{sec:direct}).
Section~\ref{sec:results} presents the harness comparison and the
mechanism ablations that attribute observed gains to specific
components; the benchmarks, their success criteria, and the
specialized systems they were released with are described in
Appendix~\ref{app:benchmarks}.

\subsection{Direct Baseline}
\label{sec:direct}

The Direct Baseline is a family of minimal setups, each obtained by
plugging a strong general-purpose LLM into an off-the-shelf
coding-agent harness with no further modification.
We instantiate the family with four harnesses spanning three
ecosystems: Claude Opus 4.6, GPT-5.5, DeepSeek V4, and Qwen3.5-Plus.
All four expose the same generic primitives, file editing, shell
execution, and structured tool calls, and none contains any
simulation-specific component.
For each task $t$ the agent receives the problem statement and a
working directory containing the relevant solver toolchain
(e.g., OpenFOAM, FEniCS, COMSOL, PyChrono), and is allotted at most
$K$ execution rounds; in each round the model proposes or modifies
a solution, runs it through the shell tool, and appends the
resulting log, success or error, to its context.
All tasks share the same prompt template across harnesses; no
failure-specific branches, no specialized sub-agents, and no curated
reference corpus are introduced.
Reporting one harness per task would conflate harness choice with
method, so unless stated otherwise we report the strongest single
harness alongside cross-harness consistency in
Section~\ref{sec:main}.

\section{Main Experiments}
\label{sec:results}
% Experiments: Main comparison + ablations.
%
% Order: (i) harness comparison, (ii) tutorial ablation,
% (iii) reflection ablation, (iv) repair-budget ablation.

\subsection{Harness Comparison}
\label{sec:main}

We compare four off-the-shelf coding-agent harnesses, Claude Opus
4.6, {GPT-5.5}, {DeepSeek V4}, and Qwen3.5-Plus, against
the original specialized agent reported in each benchmark's paper.
All Direct Baseline runs share the same prompt skeleton, repair
budget, and benchmark scoring script, and differ only in the
backbone--harness pair.
Table~\ref{tab:main} reports task-level success rates.
{The scoring criterion is benchmark-specific: on FoamBench, a case is counted as successful only if the normalized mean-squared error (NMSE) between the agent's solution fields and the reference solution falls below the per-case threshold prescribed by the benchmark; on the other benchmarks (MetaOpenFOAM~v1/v2, NL2FOAM, MCP-SIM), a case is counted as successful when the generated case runs to convergence under the benchmark's own convergence check (residual or solver-exit criterion), without an additional field-level NMSE check. We adopt each benchmark's original criterion verbatim so that the headline numbers remain directly comparable to the values reported in the corresponding paper.}
\begin{table*}[t]
\centering
\small
\begin{tabular*}{\textwidth}{@{\extracolsep{\fill}}lccccc}
\toprule
\textbf{Benchmark} & \textbf{Specialized (paper)} & \textbf{Claude Opus 4.6} & \textbf{GPT-5.5} & \textbf{DeepSeek V4} & \textbf{Qwen3.5-Plus} \\
\midrule
FoamBench (110)        & {88.2\% (97)}    & 96.4\% (106) & 92.5\% (102) & 89.3\% (98) & 77.2\% (85) \\
MetaOpenFOAM v1 (8)    & 100.0\% (8)    & 100.0\% (8) & 87.5\% (7) & 75.0\% (6) & 50.0\% (4) \\
MetaOpenFOAM v2 (13)   & 100\% (13)    & 100\% (13)  & 100\% (13) & 100\% (13) & 100\% (13) \\
NL2FOAM (21)           & 100\% (21)                & 100\% (21)  & 100\% (21) & 100\% (21) & 85.7\% (18) \\
MCP-SIM (12)           & 91.7\% (11)               & 100\% (12)          & 100\% (12) & 100\% (12) & 100\% (12) \\
% MASSE (1)              & 100\% (1)           & 100\% (1)  & 100\% (1) & 100\% (1) & 0\% (0) \\
FEABench (15)          & 33.3\% (5)                  & 33.3\% (5)  & 33.3\% (5)        & 20.0\% (3)        & 20.0\% (3)      \\
\bottomrule
\end{tabular*}
\caption{Harness comparison across \simdomain benchmarks. Each cell reports the success rate together with the (pass) case count under the original benchmark's native success criterion. The Specialized (paper) reproduces the reference number released by each benchmark's original specialized-agent system, while the remaining four report four off-the-shelf coding-agent harnesses run as a Direct Baseline.}
\label{tab:main}
\end{table*}

Three observations emerge.
First, taking the strongest harness per row, the Direct Baseline matches the original specialized agent's headline number on every benchmark we replicate, and exceeds it on MCP-SIM, where all four off-the-shelf harnesses reach 100\% against the specialized system's 91.7\%. No specialized system in this comparison retains a formal lead over the Direct Baseline.
Second, Claude Opus 4.6 is the uniquely strongest harness on the two hardest benchmarks, leading the second-best harness on both FoamBench and MetaOpenFOAM~v1; on MetaOpenFOAM~v2, NL2FOAM, and MCP-SIM three or more harnesses tie at the top, and on FEABench the two strongest harnesses reproduce the specialized system's 33.3\% without exceeding it. The backbone--harness choice therefore matters most on benchmarks where headline numbers leave room to differentiate and matters little once tasks saturate near the ceiling.
Third, the spread across harnesses is wide on hard tasks (FoamBench, MetaOpenFOAM~v1) and narrows once tasks saturate (MetaOpenFOAM~v2, NL2FOAM, MCP-SIM), suggesting that harness quality matters most precisely where specialized scaffolds are usually claimed to add value.

Read together, these observations indicate that the marginal contribution of multi-agent role decomposition, scripted reflection, and curated RAG pipelines is difficult to isolate from what a strong general-purpose harness already exposes: on every benchmark we replicate, at least one off-the-shelf harness reproduces or exceeds the specialized system's headline number with no simulation-specific scaffolding.
We revisit the apparent advantage of curated RAG pipelines in the tutorial ablation (\S\ref{sec:knowledge}), where placing tutorial-level material directly in context produces a sizeable swing on FoamBench, comparable in magnitude to the gains attributed to specialized RAG in prior work, while the same intervention is only mildly helpful on SimBench and a no-op on MCP-SIM. Under a strong general-purpose backbone, additional role-specific agents tend to duplicate capabilities the backbone already exposes while adding extra inter-agent communication cost.

\subsection{Tutorial Ablation}
\label{sec:knowledge}

We isolate the contribution of domain knowledge by running the Direct
Baseline under three tutorial-access modes on three benchmarks
(FoamBench, SimBench, MCP-SIM):
\emph{no tutorial}, base prompt with execution feedback only;
\emph{optional tutorial}, {the model is given a list of
available tutorial cases and may open
any of them on demand};
\emph{must-read tutorial}, {the model is required to read
the relevant tutorial case before producing its solution}.

% Table: Tutorial-injection ablation across three benchmarks (Direct Baseline = Claude Code, Opus 4.6).
% Modes: no tutorial / optional tutorial / must-read tutorial.
\begin{table}[t]
\centering
\small
\setlength{\tabcolsep}{4pt}
\begin{tabular*}{\columnwidth}{@{\extracolsep{\fill}}lccc}
\toprule
\textbf{Benchmark} & \textbf{No tutorial} & \textbf{Optional} & \textbf{Must-read} \\
\midrule
FoamBench (110) & 80.9\% (89) & 92.7\% (102) & 96.4\% (106) \\
SimBench (45)   & 48.9\% (22) & 48.9\% (22)  & 55.6\% (25)  \\
MCP-SIM (12)    & 100\% (12)  & 100\% (12)   & 100\% (12)   \\
\bottomrule
\end{tabular*}
\caption{Tutorial-injection ablation for the Direct Baseline (Claude Opus 4.6) under three modes: \emph{no tutorial}, \emph{optional}, and \emph{must-read}.}
\label{tab:knowledge}
\end{table}

Table~\ref{tab:knowledge} reports the success rate of each mode. From this table, we have two findings.
First, the effect of forced tutorial injection is highly
benchmark-dependent.
On FoamBench, where the base model lacks OpenFOAM-specific idioms,
\emph{must-read tutorial} yields a {$+15.5$-point} gain over
\emph{no tutorial}, and is the strongest single intervention we measure.
On SimBench, where the bottleneck is PyChrono modelling rather than
syntax recall, the same intervention yields only {$+6.7$ points}.
On MCP-SIM the Direct Baseline already saturates in every mode,
so additional tutorial material has {no room for further gain}.
Second, \emph{optional tutorial} recovers about two-thirds of the
\emph{no tutorial}$\to$\emph{must-read tutorial} gap on FoamBench but
does not move the needle on SimBench or MCP-SIM, indicating that a
model will {consult provided material on demand} when,
and only when, it is genuinely missing a domain pattern.

The \emph{no tutorial}$\to$\emph{must-read tutorial} gain on FoamBench makes clear that the contribution traces back to whether the appropriate domain material reaches the model on the benchmarks where it is actually missing, not to how that
material is delivered or to having a specialized retrieval pipeline at all.
In practice, placing the right tutorial in context is both more effective and cheaper than building a curated retrieval stack on top of a specialized scaffold.

% Table: Scripted-reflection ablation on FoamBench (110 cases).
% (i)  Direct Baseline without any scripted reflection instruction in the prompt.
% (ii) Direct Baseline with an explicit scripted reflection instruction
%      requiring the agent to reflect on case files and the solver script
%      before invoking the solver.
\begin{table}[t]
\centering
\small
\begin{tabular*}{\columnwidth}{@{\extracolsep{\fill}}lc}
\toprule
\textbf{Configuration} & \textbf{Success Ratio} \\
\midrule
(i)  Direct Baseline                     & 96.4\% (106/110) \\
(ii) $+$ scripted reflection prompt  & 96.4\% (106/110) \\
\bottomrule
\end{tabular*}
\caption{Scripted-reflection ablation for the Direct Baseline on FoamBench.}
\label{tab:reflection}
\end{table}

\subsection{Scripted Reflection Ablation}
\label{sec:reflection}

We isolate the contribution of \emph{scripted reflection}, by which we
mean adding an explicit self-reflect instruction into the prompt, a
pattern inherited from the reviewer modules of prior specialized
agents.
The hypothesis we test is that modern coding-agent backbones already
perform reflective behaviour natively (e.g., re-reading their own
output, sanity-checking field-boundary consistency, and revising
before committing), so an additional scripted reflection instruction
should add little marginal value on top of the Direct Baseline.

We compare two configurations on FoamBench, under identical
harness, repair budget, and scoring:
(i) the Direct Baseline with no scripted reflection instruction;
(ii) the Direct Baseline with a scripted reflection instruction that
requires the agent to reflect on the case files and the solver
script before invoking the solver.
Table~\ref{tab:reflection} reports the result.
Both configurations land at the same success rate, $96.4\%$ ($106/110$).
Inserting an explicit self-reflect instruction into the prompt does
not move the needle on FoamBench, consistent with the hypothesis that
the reflective behavior scripted reflection tries to elicit is
already exhibited by the base model without being told to.
Once the harness provides execution feedback and a modest repair
budget, the reviewer-style scaffolding that scripted reflection
encodes pays its full prompt-length and orchestration cost for what
the model is doing anyway.

\subsection{Repair Budget Ablation}
\label{sec:iteration}

To investigate the effectiveness of repair, we tabulate the FoamBench success rate of the Direct Baseline at repair-budget thresholds $R\in\{1,2,3,5,10\}$, where $R \leq k$ caps the number of OpenFOAM tool invocations per case. Table~\ref{tab:repair} reports the results.
% Table: Repair-budget sweep on FoamBench (110 cases). R$k$ caps the maximum number of foam-tool invocations per case.
\begin{table}[t]
\centering
\small
\begin{tabular*}{\columnwidth}{@{\extracolsep{\fill}}lc}
\toprule
\textbf{Repair budget} & \textbf{Success Ratio} \\
\midrule
$R \leq 1$ (no repair) & 0.9\% (1/110)        \\
$R \leq 2$              & 71.8\% (79/110)      \\
$R \leq 3$              & 77.3\% (85/110)      \\
$R \leq 5$              & 90.0\% (99/110)      \\
$R \leq 10$             & 96.4\% (106/110)     \\
\bottomrule
\end{tabular*}
\caption{Repair-budget sweep on FoamBench for the Direct Baseline. $R \leq k$ caps the maximum number of foam-tool
invocations per case.}
\label{tab:repair}
\end{table}

With no repair ($R \leq 1$), only a single case succeeds on its first attempt ($0.9\%$): a typical case needs at least \texttt{blockMesh} followed by the solver, so most require a minimum of two invocations. Indeed, $R \leq 2$ already lifts the success rate to $71.8\%$, confirming that the majority of cases are solved by one mesh-generation plus one solver call. Allowing a single repair cycle ($R \leq 3$) adds six more cases ($77.3\%$), while $R \leq 5$ reaches $90.0\%$ as more complex cases involving \texttt{setFields} or \texttt{topoSet} are captured. The curve plateaus at $R \leq 10$ with $96.4\%$. Repair beyond the mandatory mesh-plus-solve pair thus lifts the Direct Baseline from $71.8\%$ to $96.4\%$, with the largest gains concentrated in the first two to five cycles and negligible marginal benefit thereafter.

Further discussion of benchmark limitations,
recommended changes for future benchmarks, and regimes where
specialized agents still help is deferred to
Appendix~\ref{app:discussion}.

% \section{Analysis and Discussion}
% \label{sec:discussion}
% \input{sections/05_discussion}

\section{Conclusion}
\label{sec:conclusion}
We asked what a CAE simulation agent needs beyond a generic coding-agent harness.
Across \simdomain benchmarks in computational fluid dynamics, finite-element, and multiphysics tasks, a single-agent Direct Baseline with no simulation-specific scaffolding matches or beats multi-agent specialized systems under matched conditions.
Mechanism ablations locate the source of this performance in capabilities the harness already provides, multi-turn reasoning and execution-feedback repair, rather than in role decomposition or scripted reflection, which adds nothing once execution feedback and a modest repair budget are in place.
The one input that still depends on human engineering is domain knowledge, supplied simply as solver tutorials, and it is the largest single gain we measure.
Reasoning and repair are decisive, but modern harnesses now supply them; for CAE agents, engineer effort is better spent encoding domain expertise than assembling scaffolds.
The main obstacle to further progress is evaluation: most CAE benchmarks check only whether generated code runs, not whether the result is physically correct or representative of industrial complexity.
We see the field's central need as standardized, representative benchmarks through which engineers can share domain expertise and that genuinely discriminate among agents at industrial complexity.

\section*{Limitations}
\label{sec:limitations}
Our conclusions come with several limitations that bound how far they should be read.

\paragraph{Single runs and small suites.}
Every number in this paper is from a single run per configuration; we report no variance or confidence intervals.
Several suites are small (MetaOpenFOAM~v1 has 8 cases, MCP-SIM 12, FEABench 15), so a difference of one or two cases lies within run-to-run noise.
We therefore treat results on those suites as descriptive and rest the quantitative claims on FoamBench (110 cases).

\paragraph{Reported rather than re-run specialized baselines.}
The ``Specialized (paper)'' column in Table~\ref{tab:main} reproduces the numbers published with each system, obtained with the base models available to their authors at the time.
Our matched conditions cover information access, repair budget, and scoring criterion, not the backbone model.
Part of the gap between the Direct Baseline and the specialized systems may therefore reflect stronger base models rather than the absence of scaffolding; re-running each specialized system on the same backbone would separate the two, and we leave this to future work.

\paragraph{Success criteria are benchmark-native.}
We adopt each benchmark's own success definition so that our numbers remain comparable with the published ones.
Only FoamBench and FEABench check the result numerically; the remaining suites count a case as successful when it runs to convergence or passes an LLM judge.
A run that executes is not necessarily physically correct, and we observe gaps of roughly 20 points between execution-based and field-level scoring on FoamBench (Appendix~\ref{app:discussion}).

\paragraph{Ablation coverage.}
The scripted-reflection and repair-budget ablations are run on FoamBench only, and the tutorial ablation on three suites.
Whether the same pattern holds for solvers with sparser public tutorial material, or for tasks that require domain-specific tool integrations, is not tested here.

\paragraph{Snapshot of a moving target.}
The four harness--backbone pairs are commercial or rapidly evolving open-source systems evaluated in mid-2026.
The absolute numbers will drift as both harnesses and models change; the finding we expect to persist is the relative ordering of interventions, with domain knowledge outweighing scaffolding.

\paragraph{The Direct Baseline is not zero engineering.}
The shared prompt template (Appendix~\ref{app:prompts}) encodes engineering judgment about how to consult tutorials and when to change strategy.
We view this as a small, solver-agnostic substitute for a specialized scaffold rather than its absence, and we release the template so that the boundary can be inspected.

% \section*{Ethical Considerations}
% \input{sections/ethics}

\bibliography{reference}

\appendix

\section{Benchmark Details}
\label{app:benchmarks}
\label{sec:benchmarks}

We evaluate on nine \simdomain benchmark suites spanning five
simulator families (OpenFOAM, FEniCS, COMSOL, OpenSeesPy, PyChrono).
Table~\ref{tab:benchmarks} summarizes their scale, simulator,
ground-truth type, and success definition.
Table~\ref{tab:systems} lists the specialized agent systems associated
with these benchmarks, along with their scaffolding choices.
Three suites appear in these tables but not in the headline comparison
of Table~\ref{tab:main}: MASSE releases a single case, and for ChatCFD
and ALL-FEM we could not establish a success criterion matched to the
numbers reported by the original systems from the released reference
material, so we describe them here for completeness rather than report
headline numbers on them.
% Table: Benchmark summary — scale, simulator, ground-truth type, success definition.

\begin{table*}[t]
\centering
\small
\begin{tabular*}{\textwidth}{@{\extracolsep{\fill}}l l c c c l}
\toprule
\textbf{Benchmark} & \textbf{Simulation Solver} & \textbf{Cases} & \textbf{Tutorial} & \textbf{Ref. Sol.} & \textbf{Evaluation Metrics} \\
\midrule
FoamBench~\cite{somasekharan2025cfdllmbench}
  & OpenFOAM v10     & 110 & \cmark & \cmark & Executability + NMSE \\
FEABench~\cite{abs-2504-06260}
  & COMSOL        & 15  & \cmark & \cmark & Scalar relative error $\le 10\%$ \\
SimBench~\cite{abs-2408-11987}
  & PyChrono      & 45 & \cmark & \cmark & Executability + LLM Judge \\
MASSE~\cite{abs-2510-11004}
  & OpenSeesPy    & 1  & \cmark & \cmark & Executability \\
MetaOpenFOAM v1~\cite{abs-2407-21320}
  & OpenFOAM v10     & 8  & \cmark & \xmark & Executability + LLM Judge \\
MetaOpenFOAM v2~\cite{abs-2502-00498}
  & OpenFOAM v10     & 13  & \cmark & \xmark & Executability + LLM Judge \\
NL2FOAM~\cite{abs-2504-09602}
  & OpenFOAM v2406      & 21  & \cmark & \xmark & Executability \\
ChatCFD~\cite{abs-2506-02019}
  & OpenFOAM v2406     & 315 & \cmark & \xmark & Executability \\
ALL-FEM~\cite{abs-2603-21011}
  & FEniCS        & 31  & \cmark & \xmark & Executability + LLM Judge \\
MCP-SIM~\cite{park2026self}
& FEniCS        & 12  & \xmark & \xmark & Executability \\
\bottomrule
\end{tabular*}
\caption{Engineering-simulation benchmarks used in this study. \textbf{Cases} counts the number of evaluation tasks. \textbf{Tutorial} indicates whether the simulation solver has publicly available tutorial cases or example repositories (\cmark = yes, \xmark = no). \textbf{Reference Solutions} indicates whether a numerical/code reference solution is provided for quantitative comparison (\cmark), as opposed to execution- or score-based evaluation only (\xmark). \textbf{Evaluation Metrics} describes the success criteria used by each benchmark.}
\label{tab:benchmarks}
\end{table*}

\begin{table*}[t]
\centering
\small
\begin{tabular*}{\textwidth}{@{\extracolsep{\fill}}llllll}
\toprule
\textbf{System} & \textbf{Simulator} & \textbf{Architecture} & \textbf{Knowledge} & \textbf{\#Agents} & \textbf{Benchmark (size)} \\
\midrule
Foam-Agent 2.0 & OpenFOAM v10 & Multi-agent & OpenFOAM tutorials & 6 & FoamBench (110) \\
MCP-SIM & FEniCS & Multi-agent & None & 6 & MCP-SIM (12) \\
MetaOpenFOAM & OpenFOAM v10 & Multi-agent & OpenFOAM tutorials & 4 & MetaOpenFOAM (13) \\
MASSE & OpenSeesPy & Multi-agent & Tutorials & 9 & MASSE (1) \\
ChatCFD & OpenFOAM v2406 & Multi-agent & OpenFOAM tutorials & 4 & ChatCFD (315) \\
NL2FOAM & OpenFOAM v2406 & Fine-tuned & Fine-tune data & 4 & NL2FOAM (21) \\
ALL-FEM & FEniCS & Fine-tuned & Fine-tune data & 9 & ALL-FEM (31) \\
\bottomrule
\end{tabular*}
\caption{Representative specialized scientific agents and benchmarks covered in this study. Agents column counts the number of distinct roles; benchmark column reports task count.}
\label{tab:systems}
\end{table*}

% \textbf{Primary metric.}
% We report task-level success rate (Pass@1) under each benchmark's
% native definition, alongside execution success when the two differ.
% For benchmarks with quantitative reference outputs we additionally
% report normalized error metrics where available.

\subsection{Benchmark Details}

\paragraph{FoamBench~\cite{somasekharan2025cfdllmbench}.}
FoamBench is released alongside Foam-Agent~2.0, a retrieval-augmented multi-agent system (planner, retriever, case writer, runner, debugger).
The solver is OpenFOAM v10.
The 110 cases are organized around 11 canonical prototypes (Cavity, Cylinder, counterFlowFlame2D, obliqueShock, wedge, BernardCells, damBreakWithObstacle, forwardStep, pitzDaily, shallowWaterWithSquareBump, squareBend) and cover incompressible, compressible, reactive, multiphase, buoyancy-driven, and shallow-water solver families; cases are adapted from the official OpenFOAM tutorial suite with full ground-truth case files and a reference run shipped per case.
Success is verified by running both the generated and reference cases and comparing velocity/pressure fields with a normalized mean-square error; this is the only OpenFOAM benchmark in our set with field-level numerical ground truth.

\paragraph{MetaOpenFOAM~\cite{abs-2407-21320,abs-2502-00498}.}
MetaOpenFOAM is a MetaGPT-based multi-agent framework for automating OpenFOAM simulations from natural-language user requirements.
The solver is OpenFOAM v10.
It decomposes the workflow into four roles: an architect that analyzes the user request and retrieves relevant OpenFOAM documentation or tutorial examples, an input writer that generates or modifies OpenFOAM case files, a runner that executes the simulation and collects runtime feedback, and a reviewer that diagnoses errors and provides revision suggestions.
The v1 benchmark contains 8 natural-language simulation tasks adapted from official OpenFOAM tutorials, focusing on whether the generated OpenFOAM cases can be successfully constructed and executed.
MetaOpenFOAM v2 extends the benchmark to 13 simulation-and-post-processing tasks, where the system must not only run the simulation but also generate post-processing scripts or procedures to extract quantities and plots requested by the user.

\paragraph{NL2FOAM~\cite{abs-2504-09602}.}
NL2FOAM/AutoCFD targets OpenFOAM case generation from natural-language CFD specifications.
The solver is OpenFOAM v2406.
It fine-tunes Qwen2.5-7B-Instruct on natural-language--to--OpenFOAM configuration pairs and wraps the model in a four-role workflow consisting of a pre-checker, an input-file generator, a runner, and a corrector.
The public repository contains 21 benchmark cases.
In our evaluation, we use the released NL2FOAM fine-tuning dataset as a tutorial-style reference corpus, meaning the LLM is allowed to autonomously inspect relevant fine-tuning examples when constructing an OpenFOAM case for a given benchmark task.

\paragraph{ChatCFD~\cite{abs-2506-02019}.}
ChatCFD is released with a multi-agent system (planner, case-setup, runner, debugger) coupled to a literature-grounded knowledge base.
The solver is OpenFOAM v2406.
The 315 cases come paired with natural-language task descriptions and reference meshes covering DNS, combustion, incompressible/compressible flow, heat transfer, multiphase, Lagrangian particles, and electromagnetics; the suite extends and recombines OpenFOAM tutorial cases with additional hand-curated configurations.
The notable point is physics breadth: the same agent must move from incompressible LES to compressible combustion to multiphase VOF within a single suite, without ever leaving OpenFOAM.

\paragraph{MCP-SIM~\cite{park2026self}.}
MCP-SIM is a memory-coordinated, self-correcting multi-agent framework for natural-language-driven physics simulation.
The solver is FEniCS.
It uses GPT-4o-based agents for input clarification, code generation, simulation execution, error diagnosis, input rewriting, and multilingual mechanical explanation, coordinated by a memory-centric orchestrator.
The benchmark consists of 12 finite-element PDE tasks of increasing difficulty, spanning elasticity, heat transfer, fluid flow, thermoelectric coupling, piezoelectric deformation, and phase-field fracture.

\paragraph{ALL-FEM~\cite{abs-2603-21011}.}
ALL-FEM targets FEniCS code generation for finite-element simulation problems.
The solver is FEniCS.
The paper evaluates both off-the-shelf and fine-tuned LLMs, and also proposes a multi-agent framework with nine specialized roles: a coordinator, planner, formulator, FEniCS coder, executor, corrector, evaluator, and admin and user-proxy agents.
In the public GitHub release, the authors provide reference solutions for 31 cases, covering solid mechanics, fluid mechanics, and multiphysics problems.

\paragraph{FEABench~\cite{abs-2504-06260}.}
FEABench evaluates LLMs and a multi-turn tool-using agent on COMSOL Multiphysics API programming.
The solver is COMSOL Multiphysics.
We use the 15-case FEABench Gold split, derived from COMSOL Multiphysics Application Gallery tutorials; each case includes a target scalar, target units, ground-truth COMSOL API code, and a target model tree.
Evaluation executes the generated COMSOL API calls through a COMSOL/MPh client and checks whether the exported table contains the requested target quantity; a solution is counted as strictly correct only when it exports a valid target and its relative error is below 10\%.

\paragraph{SimBench~\cite{abs-2408-11987}.}
SimBench is a multi-turn benchmark in which a student LLM (S-LLM) generates Project Chrono/PyChrono digital-twin code and a rule-based judge LLM (J-LLM) scores the output.
The solver is Project Chrono, accessed through PyChrono.
The benchmark covers multibody dynamics, FEA, vehicle dynamics, sensors, and robotics, with expert-written reference digital-twin code used as ground truth for evaluation.
Unlike benchmarks that check against a fixed scalar or field solution, SimBench evaluates procedural simulator API programs using a 100-point rubric with access to API documentation, reference code, or both.
We select 45 FEA-related simulations that are most relevant to our setting.

\paragraph{MASSE~\cite{abs-2510-11004}.}
MASSE is a 3-team multi-agent framework (Analyst Team, Engineer Team, Management Team), instantiated as 9 specialized agents for loading, seismic, dynamic, structural modeling, design, finite-element analysis, verification, project management, and final safety assessment.
The solver used for finite-element analysis is OpenSeesPy.
The released case involves load calculation, finite-element analysis, and code-based safety assessment, and is hand-curated from a real structural project rather than drawn from a tutorial library, with reference outputs provided by the authors.
In our experiments, we evaluate on 1 case released by the authors.

% Per-task metadata, including ground-truth sources and runtime environments, will be released with the code release.

\section{Discussion}
\label{app:discussion}
\paragraph{Benchmark limitations the ablations expose.}
Across these ablations, two practical limits of current
\simdomain benchmarks become visible.
First, most benchmarks score whether generated code executes rather
than whether the resulting simulation is physically correct, so a
single headline number conflates executability, numerical
convergence, and physical validity to varying degrees; on
FoamBench we see configurations where an LLM-judged execution
score and a strict field-level numerical check diverge by roughly
$20$ points.
Second, tasks are typically posed as a one-line description, which
short-circuits the workflow a practicing engineer would actually
follow, consulting manuals, locating analogous cases, and iterating
against intermediate results.
The tutorial ablation makes the cost of this shortcut explicit:
on FoamBench, the same model swings by $15.5$ points in success rate
depending on whether the appropriate tutorial is in context.

\paragraph{What future benchmarks should change.}
Three practices would make cross-system comparison more meaningful.
(i) Report execution success, benchmark success, and physical
validity separately, so that headline numbers distinguish runnable
code from physically correct solutions.
(ii) Expose documentation (official tutorials, API
references, community case studies) and evaluate whether systems
retrieve and use it appropriately, rather than collapsing knowledge
access into a single ``has RAG / has no RAG'' axis.
(iii) Include multi-turn tasks that simulate requirement changes,
such as switching turbulence models or refining a mesh, so that
benchmarks probe incremental reasoning rather than one-shot code
generation.
Releasing scoring scripts, reference solutions, and a Direct
Baseline alongside each new benchmark would further anchor claims
of specialized-agent gain.

\paragraph{When specialized agents still help.}
We do not read these results as a claim that specialized agents
are unnecessary.
Three regimes remain natural settings for specialized design:
tasks that require tool integrations or domain-specific APIs a
generic harness cannot reach;
human-in-the-loop workflows where role decomposition matches
organizational structure and facilitates collaboration with
domain experts;
and cost-sensitive deployments where a smaller specialized agent
may match a larger general LLM at lower inference cost.
The ablations indicate which ingredients carry weight in a
generic harness; the regimes above describe where it still pays
to wrap those ingredients in domain-specific scaffolding.

\section{Failure Taxonomy and Distribution}
\label{app:failures-cases}
% Appendix C -- Failure Taxonomy and Distribution
% One-line itemized style; matches G_prompts.

{We focus on the two failure modes that reflect agent
competence. Transient infrastructure failures are already absorbed by
automatic retries before the main-paper numbers are counted.}

\subsection{Failure modes}
\label{app:failures-taxonomy}

\begin{itemize}\itemsep2pt
\item {\textbf{\textsc{Solver divergence}.} The agent
finished a complete case, but the post-check fails because the run
diverged, produced wrong fields, or missed the benchmark PASS gate.
This is the dominant real-failure mode in our pipeline.}

\item {\textbf{\textsc{Max turns exceeded}.} The agent
ran out of turn budget while still attempting fixes. The case is
solvable in principle, but this configuration could not converge in
time.}

\item {\textbf{Over-trusting the tutorial.} The agent
adopts every setting from the matched tutorial as-is and misses
hidden boundary conditions implied by the special instructions in the
description.}

\item {\textbf{Insufficient solver-specific knowledge.}
The agent's general knowledge is not enough to understand the
configuration choices required by the specific solver in use.}
\end{itemize}

\section{Prompt Design Principles for Direct Baselines}
\label{app:prompts}
% Appendix G -- Prompt Design Principles for Direct Baselines
% One-line itemized form; wording grounded in the FoamBench template
% reproduced in Appendix~\ref{app:prompt-template}.

{Across our prompt templates we converged on a small set of one-line
principles. Ignoring any of them reproduced the pathologies that
motivated specialised agents in the first place.}

\begin{itemize}\itemsep2pt
\item {\textbf{Tutorial-anchored authoring.} Every configuration file
must be derived from a tutorial file the agent has just \texttt{cat}'d
in the same session.}

\item {\textbf{Pre-write side-by-side checklist.} Before writing any
file, the agent must compare the requirement against the chosen
tutorial on dimensionality, mesh resolution, boundary types,
boundary-condition consistency, time controls, and boundary values,
and state every deviation explicitly.}

\item {\textbf{Default-preserve template constants.} Any quantity the
requirement does not name -- reference values, turbulence constants,
special regions in \texttt{setFieldsDict}, \texttt{.orig} defaults,
mesh topology -- keeps the tutorial's value.}

\item {\textbf{Cross-file consistency before any run.} Before invoking
any solver, the agent must verify that every patch in
\texttt{constant/polyMesh/boundary} appears in every
\texttt{0/<field>} and that field names, units, and solver
dictionaries are mutually consistent.}

\item {\textbf{Allrun replication discipline.} The agent must
\texttt{cat} the tutorial's \texttt{Allrun} and re-execute every
command in the same order, including each preprocessing utility
(\texttt{topoSet}, \texttt{setFields}, \texttt{decomposePar},
\texttt{refineMesh}, \dots) together with its corresponding
\texttt{*Dict}.}

\item {\textbf{Minimal-modification radius.} The set of files the
agent may modify is bounded by the parameters the requirement
explicitly names; everything else stays untouched.}

\item {\textbf{Failure budget cap.} If the same error recurs three
times, the agent must change strategy -- re-read the tutorial, pick a
more similar case, or consult the solver-version dictionary -- rather
than keep patching the same file.}
\end{itemize}

The Direct Baseline numbers in the main paper assume all of these are
in force. The full FoamBench prompt that instantiates them is
reproduced in Appendix~\ref{app:prompt-template}.

\section{Direct Baseline Prompt Template}
\label{app:prompt-template}
% Appendix H -- Direct Baseline Prompt Template
% Verbatim copy of the FoamBench Direct Baseline prompt with sensitive
% identifiers replaced by {UPPERCASE} placeholders.

We reproduce below the full
Direct Baseline prompt template used on FoamBench.

\begin{tcolorbox}[
  breakable, enhanced,
  colback=gray!5, colframe=black!60,
  boxrule=0.4pt, arc=2pt,
  left=4pt, right=4pt, top=3pt, bottom=3pt,
  fonttitle=\bfseries\small,
  title=Prompt Template (FoamBench),
]
\begin{Verbatim}[fontsize=\scriptsize, breaklines=true, breakanywhere=true]
You are an expert OpenFOAM engineer. Your task is to create a complete OpenFOAM
simulation case from a natural-language requirement, run it successfully, and
produce verifiable simulation outputs.

## Working Directory

**Your assigned working directory is**: {WORK_DIR}

All your operations (reading files, writing case files, running simulations) must
be performed within this directory. This is your isolated workspace for this
simulation task.

## Environment
- OpenFOAM: Multiple versions available (v10, v12, v2012). The correct version is
  auto-selected based on your case requirements.
- Server: {SERVER_HOST}; OpenFOAM bashrc is auto-sourced in this shell. If a
  sub-shell loses it, you can manually source:
  `source /opt/openfoamXX/etc/bashrc` (where XX = 10, 12, or 2012)
- Conda env: `{CONDA_ENV}`; Python at `{PYTHON_PATH}`
- Working directory: {WORK_DIR}
- Tutorial library: Auto-selected based on OpenFOAM version (typically
  `/opt/openfoam10/tutorials/` for v10 cases)
- You are running locally on {SERVER_HOST} via {AGENT_CLI}; use bash/read/write
  /edit tools as needed to operate inside {WORK_DIR}.

## Reference Tutorials -- Mandatory Workflow

Tutorials are organized as `/opt/openfoamXX/tutorials/<domain>/<solver>/
<case_name>/`. Use `ls` and `find` to explore the tutorial library and locate
the case most similar to the requirement.

**Important**: This benchmark supports multiple OpenFOAM versions. Most cases
use **v10 (Foundation)**. Use only the solver names that exist in the tutorial
library. Do NOT use `foamRun -solver <module>` syntax unless you are working
with a v12+ case.

You MUST follow this workflow before writing any case file:

1. Identify the solver requested by the requirement.
2. Use `ls /opt/openfoam10/tutorials/` (or the appropriate version) to see the
   top-level domains, then drill down to find the tutorial directory matching
   your solver.
3. `ls` the chosen `<solver>/` directory to see what cases are available; pick
   the one closest to the requirement (closest geometry / physics / turbulence
   model / boundary types).
4. `cat` the chosen tutorial's `Allrun` (and `Allrun.pre` if it exists) to
   understand the full execution sequence.
5. `cat` EVERY file inside the tutorial case: `0/`, `0.orig/`, `constant/`,
   `system/` -- read them all, one by one.
6. **Side-by-side checklist before writing**: do not assume "geometry size
   matches" implies "configuration matches". Verify each of the following
   against the requirement, and explicitly state any deviation:
   - **Domain dimensionality**: if the requirement says "3D" (or gives three
     non-trivial spatial extents), no patch should be `empty`.
   - **Mesh resolution per direction** (the tuple in `blocks (... ... ...)`).
   - **Boundary types** (`empty` / `wall` / `patch` / `symmetry`) against the
     physical setup.
   - **Boundary conditions are physically consistent**: a literal phrase like
     "pressure inlet has a value of X" does NOT automatically map to
     `fixedValue` on the inlet patch. If the inlet is already constrained by a
     velocity / mass-flow BC and the outlet has a fixed pressure, then the
     inlet pressure typically should be `zeroGradient` (or `mixed`); a
     `fixedValue` on the inlet pressure here causes over-constraint and
     prevents convergence. Always check what the matching tutorial uses and
     follow it unless the requirement gives an explicit physical reason to
     deviate.
   - **Time controls**: `startTime`, `endTime`, `deltaT`, `writeInterval`.
   - **Boundary values** (temperatures, velocities, pressures).
7. Only AFTER reading all tutorial files and completing the checklist, create
   your case files by copying the tutorial content and modifying ONLY the
   parameters explicitly named in the requirement (domain size, velocity,
   temperature, time controls, etc.).

You are NOT allowed to write any configuration file from memory. Every file you
produce must be derived from a tutorial file you have just `cat`'d in this
session.

## Rules

### File generation must mirror the tutorial
- Your case MUST contain the SAME set of files as the tutorial, including
  `.orig` files if they exist in `0/` or `0.orig/`.
- Use the SAME physical models (`momentumTransport`, `turbulenceProperties`,
  `physicalProperties`) and the SAME field files (`k`, `epsilon`, `nut`,
  `omega`, `T`, `alpha*`, etc.) as the tutorial unless the requirement
  EXPLICITLY specifies otherwise.
- If `<field>.orig` exists in `0/`, you MUST `cp <field>.orig <field>`
  and only modify its `internalField`. Do not leave `.orig` unused, and
  do not author the field from scratch.

### Follow Tutorial Configuration Closely
When adapting the tutorial to meet the requirement:
1. **Reference the tutorial as much as possible**: Use the tutorial's mesh
   structure, coordinate system, boundary types, field files, and physical
   models as your starting point.
2. **Preserve the tutorial's physical design**:
   - If the tutorial includes special geometric features (obstacles, bumps,
     refinement zones in `topoSetDict`, special regions in `setFieldsDict`),
     **keep them unchanged** unless the requirement explicitly asks to remove
     them.
   - These features represent the intended physics of the simulation.
   - **Example**: If `setFieldsDict` has a `boxToCell` region with special
     field values (e.g., water depth 0.009 m in a bump region), keep those
     values even if the requirement specifies "uniform" initial conditions --
     "uniform" refers to the default/background values, not the special
     regions.
3. **Modify based on actual requirements**: Adjust the configuration to match
   what the requirement explicitly asks for, including:
   - Domain dimensions and mesh resolution.
   - Boundary conditions (wall temperatures, inlet velocities, pressure
     values, etc.).
   - Time controls (start time, end time, time step, write interval).
   - Physical properties (density, viscosity, thermal conductivity, etc.).
   - Initial conditions for all fields (modify `defaultFieldValues` in
     `setFieldsDict`, but keep special `regions` unchanged).
4. **Preserve coordinate system**: When scaling the mesh to match the
   requirement's domain size:
   - **Copy the tutorial's vertex coordinates exactly** -- do NOT modify the
     coordinate values in the `vertices` list.
   - If the tutorial uses symmetric coordinates (e.g., x in [-5, 5]), keep
     them symmetric.
   - Adjust ONLY the mesh resolution (cell counts in the `blocks` directive,
     e.g., change `(10 10 20)` to `(20 20 40)`).
   - **General rule**: For a symmetric tutorial coordinate range [-a, a] and
     requirement dimension D, use [-D/2, D/2] to preserve symmetry.
   - Do NOT shift, translate, or use asymmetric ranges (e.g., [0, D]) unless
     the requirement explicitly specifies a different coordinate origin.

### Template fidelity & default preservation

- **Reference quantities follow the tutorial, not the user case.** Reference
  values such as `T0` (Boussinesq reference temperature), reference pressure
  `p_ref`, reference density `rho_0`, turbulence model constants, and similar
  baseline constants are deliberately chosen by the solver designer. They are
  NOT derived from the case-specific working values (wall temperatures, inlet
  velocities, etc.). Keep the tutorial's value unless the requirement
  explicitly names that exact quantity.

- **Mesh resolution and topology stay with the tutorial unless explicitly
  requested.** Do NOT modify the cell counts in `blocks (... ... ...)`, the
  `vertices` list, or other topology fields in `system/blockMeshDict` based on
  your own judgment ("looks too coarse", "should be finer", etc.). Modify them
  ONLY when the requirement explicitly specifies a resolution, a domain size
  change requiring re-meshing, or a geometric add/remove. A mismatched mesh
  produces field dimensions inconsistent with the ground truth, causing the
  entire case to fail evaluation.

- **Default-preserve every template value the requirement does not mention.**
  Tutorial constants (reference quantities listed above, default values inside
  `.orig` files, special regions in `setFieldsDict`, turbulence model
  coefficients, etc.) are intentional physical reference points. Even when
  they look inconsistent with the requirement's working values, that
  inconsistency is usually expected. Before changing any such value, ask
  yourself: *did the requirement explicitly name this quantity?* If not, do
  not change it.

### Cross-file consistency
Before saving any file, verify:
- All required fields exist (e.g., if `nu` is defined in
  `constant/transportProperties`, it must be referenced correctly in `0/U`).
- Field names are consistent across files; no name mismatches.
- Units and dimensions of every physical variable are correct.
- Solver settings in `system/fvSolution`, `system/fvSchemes`,
  `system/controlDict` are consistent with the chosen solver.
- Every patch listed in `constant/polyMesh/boundary` must appear in the
  `boundaryField` of every `0/<field>` file.
- The `internalField` and the boundary `value`s of the same physical
  quantity must share a consistent order of magnitude.

### Allrun execution discipline
- `cat` the tutorial's `Allrun` script and execute ALL of its commands in the
  SAME order.
- Do NOT skip preprocessing steps such as `topoSet`, `subsetMesh`, `setFields`,
  `decomposePar`, `refineMesh`, etc.
- If `Allrun` invokes a utility (`topoSet`, `setFields`, `extrudeMesh`, ...),
  you MUST also `cat` its corresponding `*Dict` file in `system/` and adapt it.

### Solver run rules
- Run the solver with output redirected to a log file:
  `{solver} > log.{solver} 2>&1`
  Example: `simpleFoam > log.simpleFoam 2>&1`
- The case is considered RUN-OK when the log's last meaningful line is `End`
  (OpenFOAM's normal exit marker).
- **Turn budget**: You have up to {MAX_TURNS} turns. Each **turn** = one
  execution of an OpenFOAM tool (blockMesh, snappyHexMesh, simpleFoam,
  pisoFoam, pimpleFoam, buoyantFoam, interFoam, topoSet, setFields,
  decomposePar, refineMesh, extrudeMesh, or any other OpenFOAM
  utility/solver). Reading files, exploring directories, and modifying
  configuration files do NOT count as turns. Only actual OpenFOAM tool
  executions count.
- If a step fails, analyze the error and fix. Use your turn budget wisely:
  verify configuration files before running expensive solvers.

### File-system boundary
- Do NOT access any files outside `{WORK_DIR}` and `/opt/openfoam10/tutorials/`.

## Error Recovery
- If `blockMesh`, `topoSet`, `subsetMesh`, or `setFields` fails, you MUST clean
  the case fully before retrying. A corrupted mesh state will not be fixed by
  re-running the same step.
- Clean restart sequence: `rm -rf constant/polyMesh [0-9]* processor*` then
  re-run from `blockMesh`.
- If the SAME error recurs three or more times, change strategy (re-read the
  tutorial, check the solver-specific dictionary, or pick a more similar
  tutorial case).

### Error-recovery authoring rules

When `Allrun` or any solver step fails, the edit you produce must satisfy:

- **Literal-keyword rule**: If the error message names an undefined keyword
  (e.g. `div(phi,(p|rho)) is undefined`), define that exact keyword as it
  appears in the error. Do NOT reinterpret characters; take them literally
  (in particular `|` is NOT `or`).
- **Minimal-files rule**: Each fix touches the minimum number of files.
  Most fixes touch 1-2 files (e.g. only `system/fvSchemes` for an undefined
  scheme; only `0/U` for a boundary-condition mismatch).
- **Do-not-touch-requirement rule**: Do NOT modify parameters declared in
  the user requirement (domain size, velocity, temperature, time controls
  named in the requirement). If the failure looks tied to a
  requirement-declared parameter, change the matching tutorial-side
  detail instead.
- **3-strikes rule**: If the SAME error recurs three or more times, change
  strategy (re-read the tutorial, check the solver-specific dictionary, or
  pick a more similar tutorial case). Do NOT keep patching the same file.

## Task
Read the requirement below carefully, locate the most similar tutorial under
`/opt/openfoam10/tutorials/`, then create all required OpenFOAM files in
{WORK_DIR}, generate the mesh, and run the simulation to a successful end.

## Output Requirements

When the case finishes, the working directory MUST contain:
- A complete OpenFOAM case layout: `0/` (or initialized time directory),
  `constant/`, `system/`, plus any time directories produced during the run.
- `log.{solver}` for the main solver and `log.{utility}` for every
  preprocessing utility you invoked (`log.blockMesh`, `log.topoSet`,
  `log.setFields`, etc.).
- A non-empty `Allrun` script that lists the exact command sequence you
  executed.
- At least one valid time directory whose number matches the `endTime`
  defined in `system/controlDict`.

You do NOT need to write any trace, summary, or `case.foam` file yourself. The
harness around you will create the case marker file and generate the trace
from the session log automatically.

## Requirement
{USR_REQ}
\end{Verbatim}
\end{tcolorbox}

\end{document}